\documentstyle[twocolumn,pre,aps,epsf]{revtex}

\newcommand{\binomial}[2]{\left(\begin{array}{c}#1\\#2\end{array}\right)}
\newcommand{\be}{\begin{equation}}
\newcommand{\ee}{\end{equation}}
\newcommand{\bea}{\begin{eqnarray}}
\newcommand{\eea}{\end{eqnarray}}
\newcommand{\ba}{\begin{array}}
\newcommand{\ea}{\end{array}}

\newcommand{\Pclass}{{\bf P}}
\newcommand{\NPclass}{{\bf NP}}
\newcommand{\NPCclass}{{\bf NPC}}
\newcommand{\pspaceclass}{{\bf PSPACE}}
\newcommand{\ksat}{{\bf K-SAT}}
\newcommand{\kcol}{{\bf K-COL}}
\newcommand{\tresat}{{\bf 3-SAT}}
\newcommand{\trecol}{{\bf 3-COL}}
\newcommand{\tvacol}{{\bf 2-COL}}
\newcommand{\half}{\frac{1}{2}}

\title{Relaxation in graph coloring and satisfiability problems}
\author{Pontus Svenson$^{(a)}$ and Mats G. Nordahl$^{(a,b)}$}
\address{tfkps,tfemn@fy.chalmers.se \\ }
\address{$^{(a)}$Institute of Theoretical Physics \\ Chalmers University of Technology and G\"{o}teborg University \\ S-412 96 G\"{o}teborg, Sweden \\ }
\address{$^{(b)}$Santa Fe Institute \\ 1399 Hyde Park Road \\
Santa Fe, New Mexico 87501 \\}
\date{January 26, 1999}

\begin{document}
\maketitle
\begin{abstract}
Using $T=0$ Monte Carlo simulation,
we study the relaxation of graph coloring (\kcol) and 
satisfiability (\ksat),
two hard problems that have recently been shown to
possess a phase transition in solvability as a
parameter is varied. 
A change from exponentially fast to
power law relaxation, and a transition to
freezing behavior are found. 
These changes take place
for smaller values of the parameter
than the solvability transition.
Results for the coloring problem for colorable and clustered
graphs and for the fraction of persistent spins for satisfiability
are also presented.
\\\\
PACS numbers: 
75.10.Hk (Classical spin models),
75.10.Nr (Spin-glass and other random models),
02.10.Eb (Combinatorics),
89.80.+h  (Computer science and technology)
\end{abstract}

\section{Introduction}

Computers have made it possible for physicists to do
experiments without leaving their offices. By simulating
systems from Nature, or simple models from
theoretical physics, a tremendous amount
of information can be gained.
The advance of computer hardware and software has made 
it possible to simulate systems 
consisting of vast numbers of interacting particles.
Physics, in turn, has inspired new algorithms for solving problems
in computer science.
For instance, the method of simulated annealing~\cite{kirkpatrickgelattvecchi},
based on Monte Carlo simulations with the Metropolis algorithm~\cite{metropolis},
can sometimes find good solutions much
quicker than traditional algorithms~\cite{johnsonaragonmcgeochschevon}.

Physicists have also started to study systems that are not found in
Nature, but instead come from computer
science~\cite{fuanderson,hogghuberman}. As an example, 
phase transitions in optimization 
problems have been discovered and studied using statistical 
mechanics~\cite{hogg:asmcsp,kirkpatrickselman}.
Other problems studied using physical methods include
the knapsack-problem~\cite{inoue:knapsack},
graph-partitioning~\cite{wongsherringtonetal}, minimax-games~\cite{minimax},
the 8-Queens problem~\cite{deanparisi}, number 
partitioning~\cite{ferreirafontanari,mertens}
and the stable marriage problem~\cite{marriage}. 
Field theory has also been used to study, e.g., 
the enumeration of Hamiltonian cycles on
graphs~\cite{higuchi} and coloring of random, planar 
graphs~\cite{eynardkristjansen,francescoeynardguitter}.

Here we present a study of the $T=0$ relaxation behavior
of hard optimization problems. Using Monte Carlo simulation,
we have measured the energy of a system starting in a random
(excited) state and slowly relaxing into the ground state or
a low-lying excited state.
We find qualitatively different
behavior for hard and easy instances of the problems.
One of many models~(e.g.,~\cite{bray:coarsening}) 
for which relaxation has been studied is
the ferromagnetic Ising model on a regular lattice. The models
studied here differ from the Ising model in several respects.
They are random, i.e., there is no pattern in
the interactions between the spins, and the interactions
have infinite range. \kcol\ can be viewed as a 
Potts model on a random graph with finite connectivity.
Another important difference is that these models
can suffer from frustration.

This article is organized in the following way:
Section~\ref{csintroduction}
introduces some concepts from computer science, while the
problems we study are described in section~\ref{theproblems}.
Previous work on the
solvability transition is reviewed in section~\ref{transitionsection}
and some approximate explanations in section~\ref{approximatetheory}.
Section~\ref{relaxationsimulations} describes our results on the
relaxation behavior, section~\ref{otherlocalsearch}
compares the results to other local search methods.
Additional measurements, e.g., of the fraction of 
persistent spins, are found in section~\ref{otherresults}.
Section~\ref{relaxationexplanation} discusses
possible explanations for the relaxation behavior 
and comments on the importance of entropy barriers.
Conclusions and a discussion are
contained in section~\ref{discussion}.

\section{Computer Science for Physicists}
\label{csintroduction}

Computer scientists classify problems
according to the maximal amount of resources needed
for their solution. The most 
important resource is time, but it is also possible to distinguish 
between problems that require qualitatively different amounts of 
memory. For example, a list of $N$ elements can always be sorted in
time less than
$k N \log{N}$, where $k$ is some constant~\cite{knuth3}.
The problems whose running time on a {\em universal Turing 
machine}~(e.g., \cite{hopcroftullman})
is bounded by a 
polynomial in their size are said to be in the class
\Pclass. The important class \NPclass\  (for 
non-deterministic polynomial) consists of those
problems where it can be checked in polynomial time
whether a proposed solution actually solves the problem.
(A {\em non-deterministic}
Turing machine would be able to solve \NPclass\ problems in 
polynomial time.)
It is obvious that $\Pclass \subseteq \NPclass$, but there is no
proof that $\Pclass \neq \NPclass$. However, most people believe that
there are \NPclass\ problems whose worst-case instances take exponential
time to solve on
a universal Turing machine.

The class \NPclass-complete (or \NPCclass) are the most important 
problems in \NPclass. A problem of size $N$ is in \NPCclass\ if all other 
\NPclass\ problems can be transformed into it in time at most
polynomial in $N$.
A method to solve an \NPCclass\ problem efficiently can thus be used
to solve any \NPclass\ problem efficiently.
It is known that if $\Pclass \neq \NPclass$ then 
there are problems in \NPclass\ that are in neither \Pclass\
nor \NPCclass. A problem is called \NPclass-hard if it is at least
as difficult as the most difficult \NPclass\ problems; \NPCclass\ is the
intersection of \NPclass\ and \NPclass-hard.

It is worth emphasizing that it is the worst case complexity that 
determines whether a problem is in \NPCclass. The average time needed to solve
an \NPCclass\ problem (given a distribution of problems)
may still be polynomial in problem size.
The properties of \kcol\ and \ksat\ studied here are related to
average case behavior, and do not address the question of whether
$\Pclass \neq \NPclass$.

A nice introduction to most elementary concepts from 
theoretical computer science 
can be found in~\cite{dewdney}. A modern reference on 
complexity theory and \NPclass\
problems is~\cite{papadimitriou}, while~\cite{gareyjohnson} has
an extensive list of \NPCclass\ problems.

\section{\ksat\  and \kcol}
\label{theproblems}

Two important problems in \NPCclass\ are graph coloring (\kcol)
and satisfiability testing ({\bf SAT}).
Graph coloring is 
the problem of
coloring a graph with $N$ vertices and $M$ edges using 
$K$ colors so that no two 
adjacent vertices have the same color. 
In physical terms,
\kcol\ is the problem of 
finding a ground state without frustrated bonds in an
antiferromagnetic $K$-state Potts model on a random graph.
Related models have been
studied by Baillie, Johnston, and coworkers 
(e.g.~\cite{bailliejohnstonkownacki,johnstonplechac}), 
who considered Potts models on $\phi^n$-model 
Feynman diagrams. They found similarities between models on $\phi^3$ and
$\phi^4$ graphs and Bethe lattices, and showed that mean-field 
theories work well for describing both ferromagnetic and
antiferromagnetic models on Feynman diagrams.

The most natural application of graph coloring is in scheduling
problems. For example, a school where each teacher
and student can be involved in several different classes 
must schedule the classes so that no collisions occur. If there
are $K$ different time slots available, this is \kcol.

Satisfiability was the first
problem shown to be in \NPCclass~\cite{cook}. It is the problem 
of finding an assignment of true or false to $N$ variables
so that a boolean formula in them is satisfied. 
In \ksat, this formula is written in {\em conjunctive normal form} (CNF),
that is, it
consists of the logical {\bf AND} of $M$ clauses, each clause being the
{\bf OR} of $K$ (possibly negated) variables, where the same clause
may appear more than once in a formula. 
For example, 
$(x \lor y) \land (y \lor \neg z)$ is an instance of {\bf 2-SAT}
with two clauses and three variables.
Applications of \ksat\ include theorem proving, VLSI design,
and learning.

In \ksat, each
clause forbids one of the $2^K$ possible assignments for its
variables. In the same way,
an edge in a graph forbids $K$ of 
the $K^2$ different colorings of its vertices.
For both problems,
there
are $M$ {\em constraints} on the solutions.
The energy $\epsilon$ of a problem instance is defined as
the number of unsatisfied constraints per variable.
Both \ksat\ 
and \kcol\ are in \Pclass\ for $K=2$ and in \NPCclass\ for 
$K\geq 3$~\cite{gareyjohnson}.
The related problem ({\bf MAX-K-SAT}) of trying to minimize 
the number of unsatisfied
clauses in \ksat\ is in \NPCclass\ even for $K=2$.

The most important problems are those
where the number of constraints is 
of the same order as the number of variables,
$M = \alpha N$. The scheduling problem described above 
fulfills this condition, for example.
For graph coloring, $\alpha = \gamma/2$, where $\gamma$ is the 
{\em connectivity} of the graph.
The connectivity (or average degree) is defined as the mean number of edges
exiting each node. For a graph with $n$ vertices and $e$ edges,
it is $2e/n$.

Below, we will use $\alpha$ for \ksat\ 
and $\gamma$ when we talk of \kcol. 
We will concentrate on \kcol; 
most results for \ksat\ are similar.

\section{The transition}
\label{transitionsection}

For physicists, an interesting property of these problems 
is that they contain phase
transitions~\cite{kirkpatrickselman,cheesemankanefskytaylor,hogghubermanwilliams}.
As the number of constraints increases, there is a transition 
from a region where almost all instances of 
the problem are solvable to
a region where practically none can be solved. 
Physically, the transition is from a region
where the ground state has zero energy to one where it is finite.
For $K=2$ the transition can be seen as a transition between a
\Pclass\ problem (finding a perfect solution to the problem) 
and an \NPCclass\ problem (finding an assignment of variables
that minimizes the number of unsatisfied constraints).

An approximation similar to mean-field theory
has been used by Williams and 
Hogg~(e.g.,~\cite{williamshogg})
and others to explain some of the properties of the phase transition.
Recently, Friedgut has also made some progress towards
showing rigorously the existence of 
a sharp transition in solvability for both
\ksat~\cite{friedgut} and \kcol~\cite{achlioptasfriedgut}.

Related to this phase transition in problem
solvability, there is a transition in
how difficult it is to solve a problem
or show that no solutions 
exist~\cite{cheesemankanefskytaylor,hoggwilliams:double,mitchellselmanlevesque}. If
there are few constraints on the solution (the problem is
{\em underconstrained}), it is easy to find one.  
Similarly, if there are so many constraints that the
problem is {\em overconstrained}, not much effort is
needed to show that it is unsolvable. 
In between these regions, where the problems are
{\em critically constrained}, there is a peak
in problem difficulty. This is called the ``easy-hard-easy'' 
transition~\cite{hoggwilliams:double}. 

\ksat\ has recently been studied by a number of physicists. 
Kirkpatrick and Selman~\cite{kirkpatrickselman}
studied the phase transition using finite-size
scaling methods, and Monasson and 
Zecchina~\cite{monassonzecchina}
used the replica method~\cite{spinglassbeyond}
to show that the entropy of \ksat\ 
stays finite at the transition. This means that below
the transition there are several solutions to each problem, all of 
which develop inconsistencies as the critical
$\alpha$ is passed.
Another problem that has been studied using
statistical mechanics is the number partitioning problem.
Mertens has recently shown that it too has
a phase transition\cite{mertens}. The relevant parameter here
is the ratio between the number of bits of input data
and the number of variables. 

Another problem in \NPCclass\ that also shows a
transition~\cite{gentwalsh:tsp} is the traveling salesperson 
problem ({\bf TSP}), where
the objective is to find a tour of minimum length visiting $N$ given
distinct cities. A difficulty in studying this problem is that
there is no natural parameter (like $\alpha$ and $\gamma$)
that distinguishes between under- and overconstrained problems.
To get one, the {\bf TSP} must be reformulated as a 
decision problem: is there a Hamiltonian path of length 
less than $l$? The parameter $l$ plays the same r\^{o}le as 
$\alpha$ --- for a given distribution of problems 
there is an $l_c$ such that 
if $l \gg l_c$, almost all instances have a
tour with length $<l$, but if $l \ll l_c$ practically no
such tours exist.
Traditionally, most \NPCclass\ problems are formulated as 
decision rather than optimization problems. 

There are many \NPCclass\ problems that contain no obvious 
parameter which makes it difficult to
say if the solvability phase transition 
exists in all \NPCclass-problems
or in just a few. There have been attempts to
formulate a more general parameter (e.g.,~\cite{constrainedness}),
with the drawback that it requires us to approximate
the number of solutions.

Phase transitions have also been found in problems
beyond \NPCclass, e.g., in {\bf QSAT}~\cite{gentwalsh:qsat}, a harder version
of satisfiability where the boolean variables are quantified by
either $\forall$ or $\exists$ (in ordinary {\bf SAT}, all variables
are existentially quantified). This problem is known to be 
\pspaceclass-complete~\cite{papadimitriou}, meaning that
it is at least as hard as all problems that can be solved
by a universal Turing machine without time limits but using
memory at most polynomial in problem size.

\section{An approximate theory for the transition}
\label{approximatetheory}

The approximation proposed by Williams and
Hogg~\cite{hogg:asmcsp,williamshogg}
assumes that the constraints in the problem are independent.
In physical terms it simply means performing 
an annealed rather than a quenched average over the disorder.
It is exact for graphs without loops and for satisfiability problems
where no variable is contained in more than one clause.
The probability that an independent constraint is
violated is $p = 1/2^K$ for
\ksat\ and $p = K/K^2 = 1/K$
for \kcol.
The probability of having none of $M$ constraints violated can
be approximated as $(1-p)^M$, ignoring correlations
between constraints, such as triangles in graphs.
The number of 
solutions for \kcol\ is then 
\be
N_{\mbox{\scriptsize sol}} = K^N (1-\frac{1}{K})^{\gamma N/2}    ,
\label{nsolhoggequation}
\ee
and for \ksat\ the expression is
\be
N_{\mbox{\scriptsize sol}} = 2^N (1-\frac{1}{K^2})^{\alpha N}        .
\ee  

Using the inclusion-exclusion principle
it is possible to write an exact
expression for $N_{\mbox{\scriptsize sol}}$~\cite{hogg:asmcsp}.
The inclusion-exclusion principle is the generalization
of the simple formula 
\[ 
P(A \cup B) = P(A) + P(B) - P(A\cap B)
\]  
from mathematical statistics. If we let $A_i$ be the event that
constraint $i$ is violated, it expresses the probability that 
any (i.e., at least one)
constraint is violated in terms of the probabilities of one, two,
three or more constraints being violated simultaneously
\be
P(\cup_i A_i) = \sum_{r=1}^{M} (-1)^{r+1} S_r    ,
\ee
where $S_r$ is the probability of exactly $r$ constraints being violated
simultaneously. The number of solutions can now be found as
\be
N_{\mbox{\scriptsize sol}} =  N_{\mbox{\scriptsize tot}} (1 - P(\cup_i A_i))           ,
\ee
where $N_{\mbox{\scriptsize tot}}$ is the number of possible assignments of
the variables, $N_{\mbox{\scriptsize tot}} = K^N$ for \kcol\ and $N_{\mbox{\scriptsize tot}} = 2^N$
for \ksat.
For \kcol, $S_1 =  M K^{-1}$, since there
are $M$ edges and each of them eliminates $K^{N-1}$ (of the $K^N$) solutions.
For $S_2$, we need to express the number of states that are
eliminated by each of two edges. This is given by 
$\binomial{M}{2} K^{-2}$, while the expression for 
\be
S_3 = \binomial{M}{3} K^{-3} + (K^{-2} - K^{-3}) t              ,
\label{threeedgesequation}
\ee
requires knowledge of the number of triangles, $t$, in the graph.
Expression~(\ref{threeedgesequation}) can be understood by noting that
if two edges in a triangle are frustrated, the third is always
frustrated too.
It can be shown that $t$ is Poisson-distributed with 
mean $\gamma^3/6$. To calculate
$S_i$ for $i\geq 4$, we also need to know the distribution of more
complex sub-graphs.

The critical value of the parameter can be approximated
as that $\gamma$ 
which gives $N_{\mbox{\scriptsize sol}}=1$ in ~(\ref{nsolhoggequation}), giving 
\be
\gamma_c = -2 \frac{\log{K}}{\log{(1-\frac{1}{K})}} .
\label{equation5}
\ee
For $K=3$, equation~(\ref{equation5}) gives 
$\gamma_c = 5.4$ for \kcol\ and $\alpha_c = 5.2$ for
\ksat. These values are larger than the experimental values of 
$\gamma_c = 4.6$ and $\alpha_c = 4.17$. 
For \ksat, this approximation has been independently introduced
several times~\cite[and references therein]{kirousis}.

This calculation of the critical value of $\gamma$ ignores 
all correlations between different constraints in the problem.
It gives an upper bound for $\gamma_c$ and is analogous to
studying a {\em forest}, a graph without cycles, in which all 
edges are violated with a probability $p$. 
Taking correlations into account
reduces the number of solutions\cite{hogg:asmcsp}.

Kirousis~et~al~\cite{kirousis}
have introduced a new method to get an upper bound for the 
critical parameter of \ksat. For $K=3$, they prove $\alpha_c \leq 4.598$,
and it is principle possible to get better bounds by including 
more terms in their expansion. A lower bound has also been
found~\cite{friezesuen}, $\alpha_c > 3.003$.
It is however difficult to 
generalize these methods to other problems, such as \kcol.

\section{The relaxation behavior}
\label{relaxationsimulations}

We have studied the relaxation of the energy $\epsilon$,
defined as the number of unsatisfied constraints per spin, 
of \ksat\ and \kcol\ using $T=0$
Monte Carlo (MC) simulations and the Metropolis 
single-flip algorithm~\cite{metropolis}. 
A simple case where the relaxation can be understood is the
ferromagnetic Ising model on a regular lattice.
For this model, the energy decreases as
$\epsilon \sim t^{-1/3}$ if the order parameter is conserved
by the dynamics, while $\epsilon \sim t^{-1/2}$
if single spin flips that allow the total magnetization to change
are used. These forms of relaxation behavior can be explained by 
noting that spins with the same
orientation will cluster and form domains~(e.g., \cite{bray:coarsening})
with a well-defined energy.
This explanation does not immediately carry over to our problems,
since there is no
known simple expression for the energy as a function of a length scale.

\begin{figure}
\centering
\leavevmode
\epsfbox{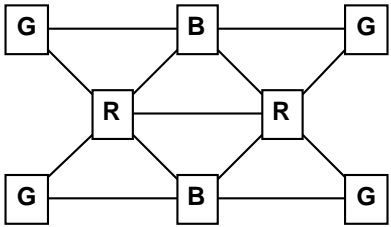}
\vspace*{2mm}
\caption{This graph is 3-colorable, but the Monte Carlo
algorithm can get stuck in the local minimum shown (letters denote
colors).}
\label{smallgraphfigur}
\end{figure}

For the \NPCclass\ problems studied here,
the energy can not be expected always to reach zero, since
there are no perfect solutions to problems with many constraints.
The $T=0$ MC algorithm can also get stuck in local
minima. A small graph where this can happen for \trecol\ is shown
in figure~\ref{smallgraphfigur}.

In each time step in our simulations $N$ spin flips are attempted. 
Each flip consists of selecting a random spin and
randomly changing its color. 
The flip is allowed
if this leaves the energy unchanged or lowers
it, otherwise the spin is left unchanged.
In temperature $T>0$ simulations, a flip that raises the
energy $\Delta$ units is allowed with probability $\exp[-\Delta/T]$.

For \kcol, we generated the problems by randomly selecting $M$
distinct edges from the $\binomial{N}{2}$ possible with $N$
vertices. (In graph terminology~\cite{bollobas}, this corresponds
to using the ${\cal G}(N,M)$ model for random graphs.) For \ksat,
each clause was generated by selecting $K$ variables, where each variable 
was negated with probability $\half$.
The formula was then generated by performing
this process $M$ times. Clauses with repeated variables
were allowed in the expressions, and also repetition of clauses.

In all random systems, the question of whether or not
the measured quantities are {\em self-averaging} is important.
Self-averaging means that the properties of the entire (infinite)
system can be understood in terms of the properties of its local
subsystems.
For random systems with local interactions, there is
a simple argument for this~(see e.g., \cite{binderyoung}), but
if the interactions are global, the situation is more complex.
However, for spin glasses such as the Sherrington-Kirkpatrick
model, the energy and other simple quantities are 
self-averaging, and we
assume that the energy is self-averaging also in \ksat\ and
\kcol. This assumption of self-averaging 
is supported by our simulations;
averaging over a small number of large graphs appears to give
the same results as averaging over many small graphs.
Schreiber 
and Martin~\cite{schreibermartin} have recently studied
various local search methods for the graph partitioning
problem and found strong numerical evidence for
self-averaging even for sparse graphs. They also provide
some arguments for why self-averaging should hold, and
conjecture that their result holds for all constraint satisfaction
problems.

Our simulations show a transition between qualitatively different 
forms of relaxation behavior for \ksat\ and \kcol. 
For small values of $\gamma$ or $\alpha$, we find exponential
relaxation to zero energy. For larger values of the parameter,
the relaxation becomes algebraic.
For large enough $\gamma$ or $\alpha$, the energy freezes 
at a non-zero value.

The change between exponential and power law relaxation occurs for
smaller values of $\gamma$ and $\alpha$ than the transition 
in solvability.
More extensive numerical investigations will be necessary
to determine the exact nature of this transition, in particular
whether it is a sharp transition, or whether there are intermediate
regions with different forms of relaxation behavior. 
In section~\ref{otherlocalsearch} below, we study 
the freezing transition, where the final energy of the MC
algorithm changes from zero to a non-zero value. When this
quantity is measured, a sharp transition is observed. This
transition does not necessarily coincide with the change in
functional form of the decay. 

\begin{figure}
\centering
\leavevmode
\epsfxsize = .75 \columnwidth
\epsfbox{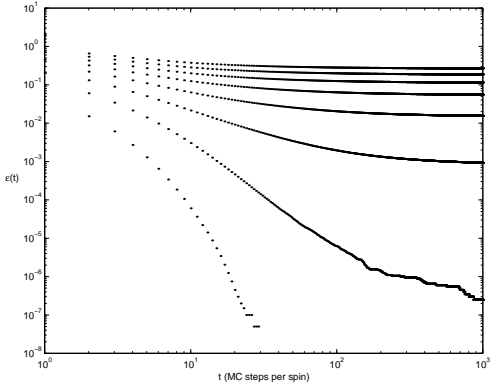}
\vspace*{1mm}
\caption{The energy per spin as a function of time averaged over two
graphs of size $10^7$ for \trecol. From bottom left to top right, 
$\gamma=1,2,3,4,5,6,7,8$.}
\label{energycolfigur}
\end{figure}

\begin{figure}
\centering
\leavevmode
\epsfxsize = .75 \columnwidth
\epsfbox{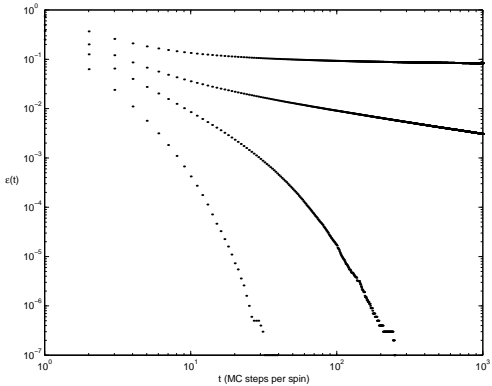}
\vspace*{2mm}
\caption{The energy per spin as a function of time for \tresat\ with
$10^6$ variables, averaged over 10 different formulas.
From bottom left to top right, $\alpha = 2,3,4,6$.}
\label{loglogenergysatfigur}
\end{figure}

Figure~\ref{energycolfigur} 
shows the relaxation behavior for \trecol\ in a log-log diagram,
while figure~\ref{loglogenergysatfigur}
shows that the behavior for \tresat\ is similar.
The data in these figures was determined 
for systems of size $10^7$ (for \trecol) and $10^6$ (for \tresat).
The relaxation showed similar behavior for other values
of $K$ as well.

\begin{figure}
\centering
\leavevmode
\epsfxsize = .75 \columnwidth
\epsfbox{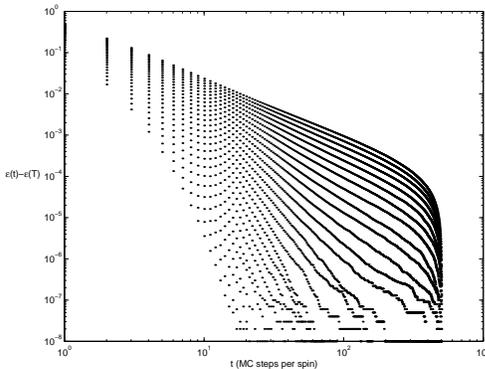}
\caption{The energy per spin as a function of time, with the
energy at 500 MC steps subtracted, for \trecol\
with $\gamma=0.5,0.6,\ldots 3.0$. 
The figure illustrates the
transition from fast to power law relaxation.}
\label{exppowerlawtransitionfigur}
\end{figure}

\begin{figure}
\centering
\leavevmode
\epsfxsize = .75 \columnwidth
\epsfbox{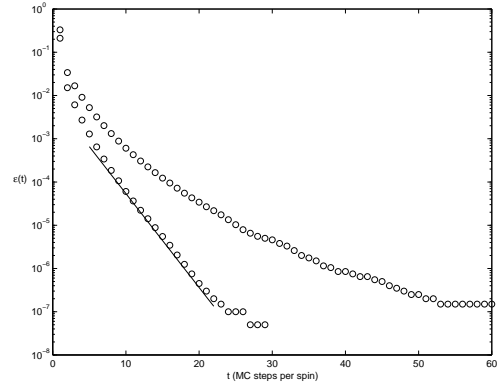}
\vspace*{2mm}
\caption{Connectivity $\gamma=1$ (lower curve) in \trecol\
gives a reasonable fit to an exponential $\exp{(-t/2)}$ for $5 \leq t \leq
22$.
Data for $\gamma=1.5$ (upper curve) is also shown.}
\label{expsmallfigur}
\end{figure}

The change between exponential and power law decay is
illustrated in more detail for \trecol\ 
in figure~\ref{exppowerlawtransitionfigur},
where we have plotted $\epsilon(t)-\epsilon(500)$ for systems of
$N=10^6$ spins. The figure shows data for $\gamma=0.5$ up to 3.0 
in increments of 0.1.
For small $\gamma$, the decay is exponential (see also 
figure~\ref{expsmallfigur} below for $\gamma=1$). When $\gamma$ is 
increased, 
a change from exponential to power law behavior is observed. 
A reasonable fit to power law behavior is found approximately
for $\gamma \geq 2$.

\begin{figure}
\centering
\leavevmode
\epsfxsize = .75 \columnwidth
\epsfbox{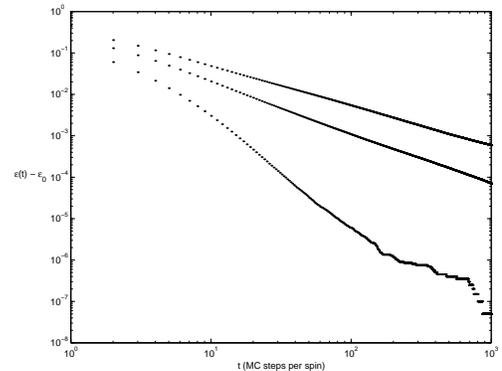}
\vspace*{2mm}
\caption{For \trecol, subtracting a constant from the energy
gives power law relaxation for $\gamma = 2$ (bottom curve),
$3$ (middle curve) and $4$ (top curve).
Note the finite-size 
effects in the bottom curve.}
\label{powerlawfigur1}
\end{figure}

\begin{figure}
\centering
\leavevmode
\epsfxsize = .75 \columnwidth
\epsfbox{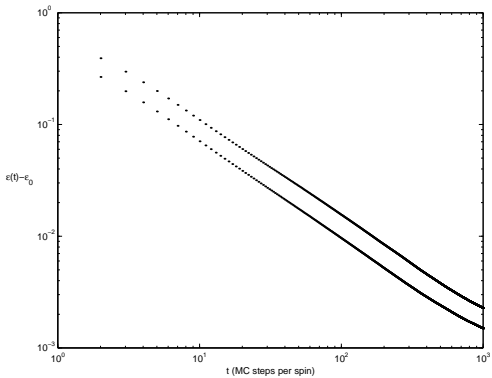}
\vspace*{2mm}
\caption{For \trecol, subtracting a constant from the energy 
gives power law relaxation with approximately identical exponents 
$\mu \approx 0.85$ for $\gamma = 5$, $6$, $7$ and $8$.
Data is plotted for $\gamma = 5$ (lower curve)
and $\gamma = 8$ (upper curve).}
\label{powerlawfigur2}
\end{figure}

Figure~\ref{expsmallfigur} shows the data for $\gamma = 1$, where
a reasonable fit to exponential relaxation is obtained, 
and for $\gamma = 1.5$,
where a cross-over behavior is seen (these and the following
data were obtained for systems of size $N=10^7$)
Figure~\ref{powerlawfigur1}
shows the data for $\gamma = 2$, 3, and 4, where a power law 
$\epsilon \sim \epsilon_0 \; + \; t^{-\mu}$
is found. Figure~\ref{powerlawfigur2} shows that the power law
also applies for $\gamma=5$ and 8; here the exponent
is given by $\mu \approx 0.85$.

The exponents for \trecol\
are summarized in
table~\ref{colrelaxtable}; note that due to 
finite-size effects only
data up to $t=200$ was used to determine the exponent for
$\gamma =2$.

\begin{figure}
\centering
\leavevmode
\epsfxsize=.75 \columnwidth
\epsfbox{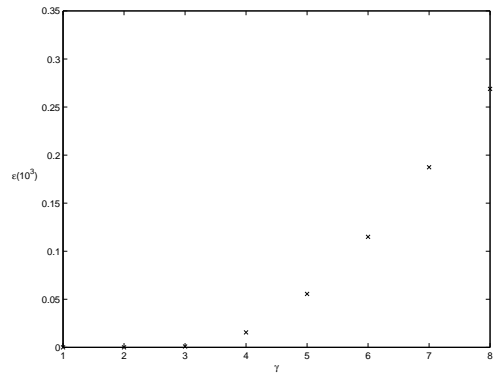}
\vspace*{3mm}
\caption{The value at which the energy freezes
as a function of $\gamma$ for
\trecol\ with $10^7$ variables.}
\label{slutenergifigur}
\end{figure}

The values for which the energy freezes 
for different $\gamma$ in \trecol\ are shown in 
figure~\ref{slutenergifigur}. For large $\gamma$,
an approximately linear increase is observed.

\begin{figure}
\centering
\leavevmode
\epsfxsize=.75 \columnwidth
\epsfbox{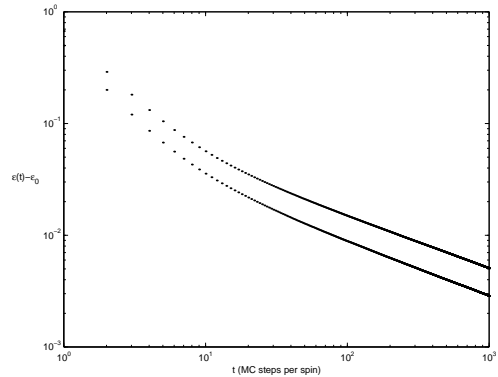}
\vspace*{2mm}
\caption{For \tresat, power law relaxation is
obtained for $\alpha=4$ (bottom curve) and 6 (upper curve).
The same data is shown as in figure~\ref{loglogenergysatfigur}, but with an 
$\alpha$-dependent constant subtracted from the energy.}
\label{tresatexponentfigur}
\end{figure}

For \tresat, $\alpha=2$ gives exponentially fast decay, 
while there appears to be a crossover behavior for
$\alpha=3$. For $\alpha =4$ and 6, power law relaxation
is obtained, 
$\epsilon \sim  {\bf \epsilon_0} \; + \; t^{-\mu}$ with 
$\mu \approx 0.6$ in both
cases, see figure~\ref{tresatexponentfigur}. As in \trecol,
$\epsilon_0$ was found to increase approximately 
linearly with $\alpha$.

No significant change in the behavior was
seen in finite temperature simulations.
Raising the temperature makes it possible to escape from
one local minimum, but the system is then trapped in another before
the ground state is reached. Raising the temperature further
repeats this scenario but also increases the fluctuations in
the energy. For high enough temperatures, the fluctuations take
over completely and the system is not trapped in any local minimum.

\begin{table}
\centering
\leavevmode
\begin{tabular}{|cccccccccc|}
$\gamma$ &  2    & 3    & 4    & 5    & 6     & 7     & 8   \hspace*{-2mm} \\
\hline
$\mu$    & 2.7   &  1.3 & 1.0  & 0.85 & 0.85  & 0.85  & 0.85 \hspace*{-2mm} \\
\end{tabular}
\vspace*{5mm}
\caption{Approximate exponents for \trecol. The relaxation
behavior can be described by 
$\epsilon \sim \epsilon_0 + t^{-\mu}$, with different 
constants $\epsilon_0$ for different $\gamma$'s. The exponents
were determined
using two graphs of size $10^7$.}
\label{colrelaxtable}
\end{table}

\begin{figure}
\centering
\leavevmode
\epsfxsize=.75 \columnwidth
\epsfbox{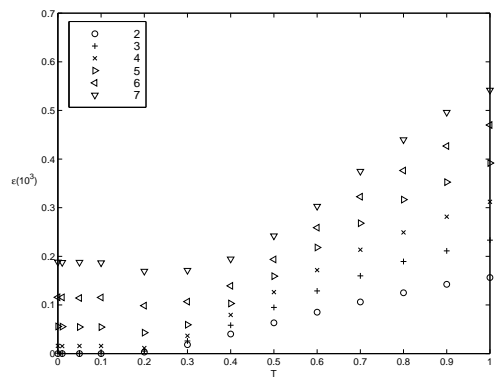}
\vspace*{3mm}
\caption{The energy after $10^3$ MC updates per spin as a function
of temperature for $\gamma =2$ to 7. }
\label{frystempberoendefigur}
\end{figure}

We found the same form of power law relaxation with
approximately identical exponents
for temperatures up to $T=0.4$, but the frozen-in value of the
energy, $\epsilon_0$, depended on the temperature,
see figure~\ref{frystempberoendefigur}. For $\gamma = 4$,
the $T=0.2$ runs were able to achieve an almost 30\% better state
than the $T=0$ runs.

Using simulated annealing to find the ground states of some
\NPclass-complete spin glasses 
($3D$ $\pm J$ and infinite range models),
Grest~et~al~\cite{grestsoukoulislevin} found a
logarithmic decay of the energy, 
$\epsilon \sim \epsilon_0 + 1/\log{t}$. Similar decay was
also recently found by K\"{u}hn~et~al~\cite{kuhnlinpoppel},
who used an algorithm where attempts were made to flip
several spins at once. The number of simultaneous flips,
which plays the same r\^{o}le as the 
temperature in simulated annealing, was
then slowly decreased. These methods will always find 
the ground state, whereas the $T=0$ MC algorithm studied
here can get stuck in local minima. The faster relaxation 
of the $T=0$ MC method thus
comes at the price of having no guarantee of finding the
ground state.

The exact values of the critical parameter for \kcol\
varies depending on the
ensemble of graphs used~\cite{hogg:asmcsp}. We
tried different ensembles and found no significant
differences in the relaxation behavior 
(e.g., $\phi^{n}$-model Feynman diagrams show approximately
the same behavior as random graphs with connectivity $\gamma=n$).

Since the MC algorithm can get stuck in local minima, one should 
verify that the behavior does not depend on the initial
values of the spins. It is also important to check to what 
extent the result depends on the choice of random graph. 
To test this, we have performed
simulations where in addition to averaging over several different
graphs we also restarted the MC algorithm with different inital
spin configurations. In particular, \trecol\ with $\gamma =2$ and 4
was studied in detail.

\begin{figure}
\centering
\leavevmode
\epsfxsize=.75 \columnwidth
\epsfbox{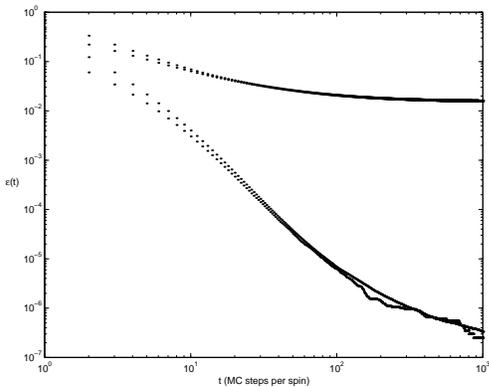}
\caption{For \trecol\
the mean energies of several runs with $(N_g,N_r)$ = (1,1000)
are compared to the results for $N=10^7$ from above for 
$\gamma=2$ (lower curves, 56 different runs) and
$\gamma=4$ (upper curves, 54 different runs).}
\label{3colgamma24_comparison}
\end{figure}

In figure~\ref{3colgamma24_comparison}
we compare the results for $N=10^7$ shown above to runs for a smaller
system ($N=10^4$), where an average over 1000 initial states was
performed for each of a larger set of graphs. The figure shows
the average energy from these runs and  $\epsilon(t)$
from the runs with $N=10^7$ for $\gamma=2$ and 4. A reasonable
agreement is found in both cases.

\begin{figure}
\centering
\leavevmode
\epsfxsize=.75 \columnwidth
\epsfbox{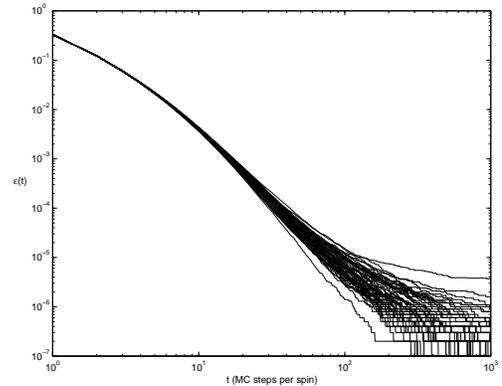}
\caption{The energy as a function of time for 56 runs
with $(N_g,N_r)$ = (1,1000), for \trecol\ with $\gamma=2$.}
\label{ng1000_g2_energy}
\end{figure}

\begin{figure}
\centering
\leavevmode
\epsfxsize=.75 \columnwidth
\epsfbox{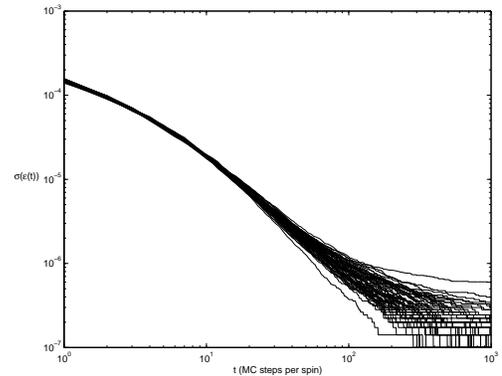}
\caption{The standard error as a function of time for 56 runs
with $(N_g,N_r)$ = (1,1000), for \trecol\ with $\gamma=2$.}
\label{ng1000_g2_stderr}
\end{figure}

\begin{figure}
\centering
\leavevmode
\epsfxsize=.75 \columnwidth
\epsfbox{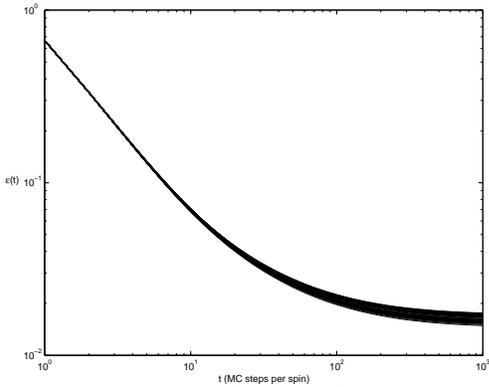}
\caption{The energy as a function of time for 54 runs
with $(N_g,N_r)$ = (1,1000), for \trecol\ with $\gamma=4$.}
\label{ng1000_g4_energy}
\end{figure}

\begin{figure}
\centering
\leavevmode
\epsfxsize=.75 \columnwidth
\epsfbox{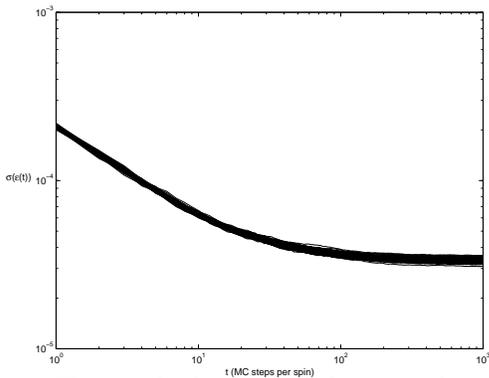}
\caption{The standard error as a function of time for 54 runs
with $(N_g,N_r)$ = (1,1000), for \trecol\ with $\gamma=4$.}
\label{ng1000_g4_stderr}
\end{figure}

The variation of the result depending on the choice of graph and 
initial state is further illustrated in figures~\ref{ng1000_g2_energy}
and~\ref{ng1000_g2_stderr}, 
which show the average energy and the standard error 
\[
\sigma = \frac{\sqrt{\langle \epsilon^2 \rangle - 
\langle \epsilon \rangle ^2 }}{\sqrt{N_g N_r}}   ,
\]
(where $\langle \cdot \rangle$ denotes the average over graphs and
restarts, and $N_g$ and $N_r$ stand for the number of different
graphs and the number of runs, respectively)
for $N=10^4$ and $\gamma=2$, for each of 56 different runs 
with $N_r=1$ and $N_g=1000$.
Figure~\ref{ng1000_g4_energy} shows
the energy and standard error for
$N=10^4$ and $\gamma=4$ for 54 different runs with $N_r=1$ and
$N_g=1000$, while figure~\ref{ng1000_g4_stderr} shows the 
standard errors from each of these runs. 

For $\gamma=4$, the variation among runs on a single graph
(see figure~\ref{ng1000_g4_stderr}) 
is very small compared to the variation among randomly 
generated graphs shown in figure~\ref{ng1000_g4_energy}; 
for $\gamma=2$ these are of comparable magnitude.

We also found similar relaxation behavior
when we started with all spins having the same value, and
did not find any differences using different
random number generators. The generators used include 
the Mitchell-Moore additive generator~(e.g.,~\cite{knuth2}),
a 48-bit multiplicative congruential generator with
multiplicative and additive constants 25214903917 and 11,
respectively, and the standard C library {\tt rand()} function.

\section{Comparison to other local search methods}
\label{otherlocalsearch}

The Monte Carlo method is an example of a {\em local search} method.
Local search methods start with a candidate solution and try to
improve it by changing it locally (e.g., by flipping a spin).
Various other such methods have been used to
study the phase transition in search cost 
in \NPCclass-problems. Gent and Walsh~\cite{gentwalsh93} studied
small problems using the {\bf GSAT} method, 
a hill-climbing procedure for solving \ksat\ problems,
and found exponential relaxation over very small time ranges.
The {\bf GSAT} method is similar to MC simulations, but
differs in the way a spin is chosen to flip. Instead of
flipping a random spin, {\bf GSAT} selects that spin whose
flipping will decrease the energy the most. 
Clark~et~al~\cite{clarketal} used only
solvable problems (determined by first using a complete
backtracking method) and found an easy-hard-easy transition in
search cost for two local search methods. 
The hardest problems occurred approximately at the true transition.

\begin{figure}
\centering
\leavevmode
\epsfxsize = .75 \columnwidth
\epsfbox{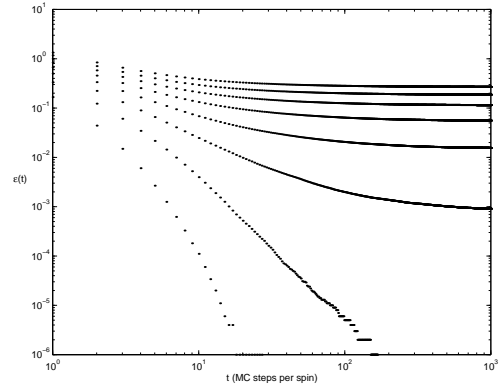}
\vspace*{2mm}
\caption{The relaxation in \trecol\
with only colorable graphs allowed. No significant change
compared to figure~\ref{energycolfigur} is observed. From bottom to top, 
$\gamma =1$, 2, 3, 4, 5, 6, 7 and 8.}
\label{k3solvcolplot}
\end{figure}

\begin{figure}
\centering
\leavevmode
\epsfxsize = .75 \columnwidth
\epsfbox{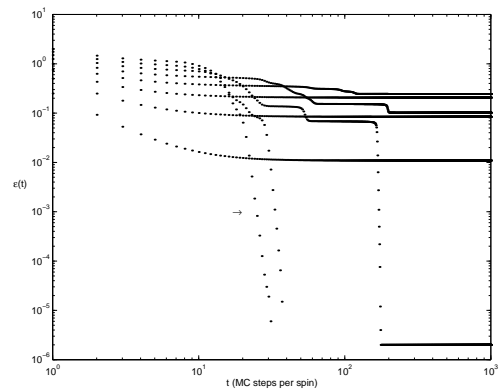}
\vspace*{2mm}
\caption{The relaxation in \tvacol\
with only colorable graphs allowed. The behavior is
similar to that found for ferromagnetic random graphs.
$\gamma=1$ starts with the lowest energy, followed by 
$\gamma=2$,3,4,5,6,7,8. The small arrow indicates the
$\gamma=8$ data.}
\label{k2solvcolenergyfigur}
\end{figure}

In our case, the behavior is quite different. 
This is particularly evident
in figure~\ref{k3solvcolplot}, which shows the relaxation
for \trecol\ with $10^5$ spins using only colorable 
graphs. It shows the same behavior as in figure~\ref{energycolfigur} 
for both colorable and uncolorable problems. 
An exceptional case is \tvacol, where we
found different behavior when only colorable graphs were used,
see figure~\ref{k2solvcolenergyfigur}. The reason for this is
that for $K=2$ and colorable graphs there is no difference between
the ferromagnetic and antiferromagnetic models. Results and 
arguments for the relaxation behavior of ferromagnetic Potts models
on random graphs will be presented elsewhere~\cite{unpublished}.

Problems with large connectivities are thus always hard to solve
using the Monte Carlo method.
This is a short-coming of the MC algorithm ---
even if the problem is solvable, the MC algorithm can get 
stuck in local minima that the other, smarter local search methods
manage to avoid.

In order to quantify the difference between the MC algorithm and other
local search methods, and to compare the freezing transition 
with the solvability transition, we determined the fraction $\eta(t)$ of
problems for which the MC method did not find an $\epsilon=0$
ground state in less than $t$ MC steps per spin.
We found that $\eta$ displayed a behavior similar to the
fraction of solvable problems --- there appears to be
a phase transition, but for a smaller value of $\gamma$.

\begin{figure}
\centering
\leavevmode
\epsfxsize = .75 \columnwidth
\epsfbox{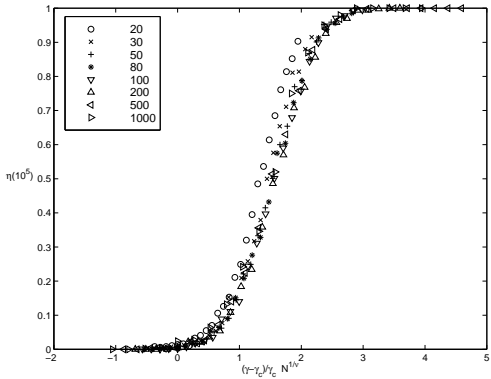}
\vspace*{4mm}
\caption{Finite-size scaling analysis of the fraction of problems 
for \trecol\
where a zero energy ground state has not been found
before $10^5$ MC updates per spin. The system size $N$
ranges from
20 to 1000, and the data is plotted against the rescaled parameter 
$(\gamma/\gamma_c-1) N^{1/\nu}$, with $\gamma_c=2.4$ and $\nu=3.75$.
A good fit is obtained, except for $N=20$ and 30.}
\label{k3coltidfss}
\end{figure}

In figure~\ref{k3coltidfss}, we plot $\eta(10^5)$ for \trecol\
against the rescaled parameter
\be
(\frac{\gamma}{\gamma_c} -1 ) N^{1/\nu}
\ee
with $\gamma_c = 2.4$ and $\nu = 3.75$ and $N$ ranging from 20 to 1000.
It is clear that 
there is a freezing transition well below the occurrence of
the solvability transition.
However, it is necessary to 
be cautious when drawing conclusions from small systems.
The largest systems we have simulated are of size $10^7$ for \trecol\ and
$10^6$ for \tresat; these systems were used to fit the
relaxation behavior shown in table~\ref{colrelaxtable}
above. To determine the freezing transition,
we used considerably smaller systems.

The approximate values where the transition in 
figure~\ref{k3coltidfss} was
found are shown in table~\ref{valuetable} for
$2 \leq K \leq 5$, together with
the experimental values for the
solvability transition for \ksat\ from~\cite{kirkpatrickselman},
and values for the solvability transition for \kcol. 
The latter were obtained by ourselves using a
non-optimized backtrack-search program with the 
Brelaz heuristic~\cite{brelaz}. 
For $K=3$, our value
coincides with the literature~\cite{hogg:asmcsp};
we are not aware of any values for $K=4$ and $5$
in the literature.
For $K\geq 4$, the values for the solvability transition are
probably not very accurate --- the results vary strongly with 
the number of variables and the number of graphs tested.
It is clear, however, that
the relaxation transition happens for smaller values
than the solvability transition. 

\begin{figure}
\centering
\leavevmode
\epsfxsize = .75 \columnwidth
\epsfbox{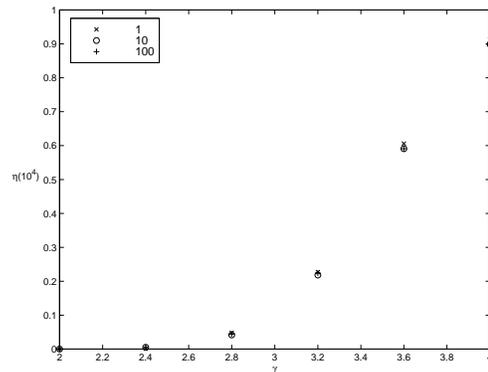}
\caption{The fraction $\eta$ of problems 
for \trecol\ where a zero energy ground state has not been found
after $10^4$ MC updates per spin, averaged over 1000 different graphs
of size 100, with 1, 10 or 100 restarts per graph.}
\label{etamultiplerunsfigur}
\end{figure}

Most of our data for the freezing transition was determined by
averaging over between 100 (for $N=1000$) and 1000 different graphs with
a single MC run on each graph. We have also made some runs where
we restarted the MC algorithm using different initial spin
configurations for each graph. For small $N$, $\eta(t)$ varied about
13\% depending on whether we used 1, 10 or 100 restarts, but for
$N\geq 80$ there was essentially no difference between the different
runs, see figure~\ref{etamultiplerunsfigur}. In each of these runs 
we averaged over 1000 different graphs.

The precise location of the freezing transition depends on the
algorithm used to determine $\eta(t)$. For example,
finite temperature MC methods give a higher value for the location
of the transition. This illustrates the difference between the
freezing transition and the solvability transition. The former is
a property of the method used to solve a problem, while the latter
is a characteristic of the problem itself.


\begin{table}
\centering
\leavevmode
\begin{tabular}{|c|c|c|c|c|}
\hline
  $ K =$         &  2      &   3   &  4       &  5           \\
\hline
\kcol\           & 0.2     & 2.4      & 6.7     & 11     \\
$\gamma_c^{exp}$ &         &  4.6     & 8.7  & 13.1      \\
\hline 
\ksat\           & 1       & 4        &  8.4     & 16.7        \\
$\alpha_c^{exp}$ & 1.0     & 4.17     & 9.75     & 20.9         \\
\hline
\end{tabular}
\vspace*{6mm}
\caption{The approximate values of
$\gamma$ and $\alpha$ where the freezing
transition was found, determined using a simple FSS scaling.
Also shown are
experimental values $\gamma_c^{exp}$ and $\alpha_c^{exp}$
for the solvability transition
determined by Kirkpatrick and
Selman and by us.}
\label{valuetable}
\end{table}

\section{Other results}
\label{otherresults}

We have
also found interesting scaling relations for other quantities.
For graph coloring, the energy as we have defined it above measures 
the fraction of frustrated edges. Another possibility is to measure
the fraction of frustrated spins, $\epsilon_s$. This quantity 
shows the same relaxation behavior, if used in the 
Metropolis algorithm instead of $\epsilon$.

\begin{figure}
\centering
\leavevmode
\epsfxsize = .75 \columnwidth
\epsfbox{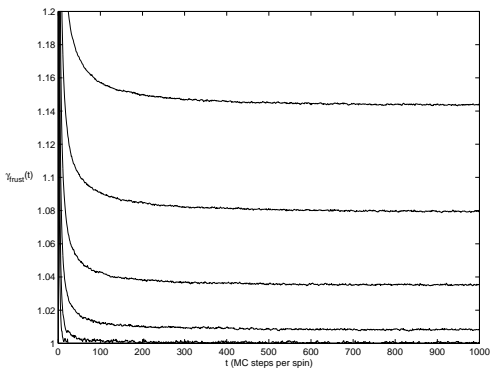}
\vspace*{2mm}
\caption{The connectivity of the graph spanned by the
frustrated edges as a function of time. 
Note the change from
$\gamma_{\mbox{\scriptsize frust}} =1$ to $>1$ as the problem difficulty increases.
The data are for \trecol\ with $N=10^5$, averaged over
10 graphs.
From bottom to top, $\gamma=2,3,4,5,6,7$.}
\label{frustratedconnectivityfigur}
\end{figure}

It is
also interesting to study the ratio 
$\gamma_{\mbox{\scriptsize frust}} = \frac{2\epsilon}{\epsilon_s}$, which is the connectivity of 
the sub-graph spanned by the frustrated edges. 
Figure~\ref{frustratedconnectivityfigur} suggests
a transition from $\gamma_{\mbox{\scriptsize frust}} = 1$ for easy
problems to $\gamma_{\mbox{\scriptsize frust}}>1$ for harder. 
If the connectivity is 1, all frustrated edges are isolated,
while connectivity 2 would mean that
all frustrated spins were connected in chains. 

\begin{figure}
\centering
\leavevmode
\epsfxsize = .75 \columnwidth
\epsfbox{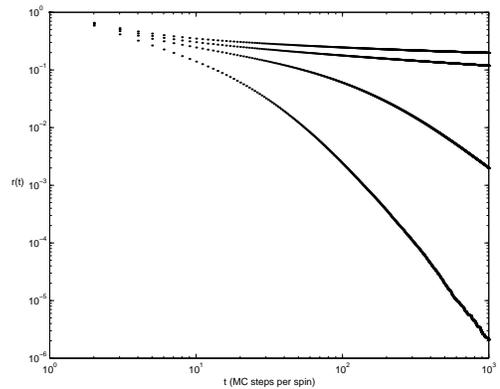}
\vspace*{3mm}
\caption{For \tresat, the fraction of persistent spins
shows a similar transition as the energy.
The data are averages of 10 runs on systems
of $10^6$ spins and show $\alpha=2$ (bottom), 3, 4, and 6 (top).}
\label{satnotflipfigur}
\end{figure}

The Monte Carlo
dynamics itself has interesting properties.
Figure~\ref{satnotflipfigur} shows the fraction $r(t)$
of persistent spins,
i.e., those that have
not yet been flipped, as a function of time for \tresat.
The data suggest a transition 
from an exponentially fast to a logarithmically slow decay as
$\alpha$ is increased (this will be explored further elsewhere).
For the Ising and Potts models on a square lattice, this
quantity has been found to scale with a power law dependence
on time~\cite{brayderridagodreche,derrida}. 

\section{Heuristic arguments for power law relaxation}
\label{relaxationexplanation}

In \kcol\ and \ksat\ there are global barriers to local 
improvement. In order to lower the frustration for a spin,
we have to make changes to many other, unfrustrated
spins (compare figure~\ref{smallgraphfigur}). One way of
explaining the relaxation is to look at these
barriers and see how long it takes to overcome then.

In the following, we call a proposed assignment of the $N$ 
variables a state in the space of all solutions.
Consider the probability $p(i)$ that a state has $i$ unsatisfied
constraints, so that $\epsilon=i/N$. 
In analogy with expression~(\ref{nsolhoggequation}) for 
\kcol, this can be approximated by 
\be
p_{\mbox{\scriptsize sol}}(i) = (1-p_v)^{M-i} p_v^i \binomial{M}{i}        ,
\ee
where $p_v$ is the probability that a constraint is violated.
The binomial factor $\binomial{M}{i}$ represents
the number of different
ways to choose the $i$ unsatisfied edges. Recall that
for \kcol\ $p_v = 1/K$ and for \ksat\ $p_v = 1/K^2$.
In the approximation,
we neglect all correlations between the 
constraints, such as the presence of triangles and other
regular structures in the graph to be colored.
In the following, only \kcol\ will be considered.

The Monte Carlo algorithm works by generating a new state,
and changing to this state if its energy is lower
than or equal to that of the current state.
In the approximation we assume that
the probability that the new state has energy $e'$ is
proportional to $p_{\mbox{\scriptsize sol}}(e')$. Given a state with energy $e$,
all transitions to states with energy $e' > e$ are forbidden. 
The probability of staying in 
a state with energy $e$ is then  
proportional to $\sum_{i=e}^{M} p_{\mbox{\scriptsize sol}}(i)$,
while the probability for a transition to an energy $e'' < e$
is proportional to $p_{\mbox{\scriptsize sol}}(e'')$. 

This process can be viewed as a Markov chain. The transition matrix
between states of different energy has elements $p_{ij}$ 
given by the probability that state $i$ is followed by state $j$:
\be
p_{ij} = 
\left\{
\begin{array}{cr} 
0, & i < j\\
\sum_{n=i}^{M} p_{\mbox{\scriptsize sol}}(n) =   
1-\sum_{n=0}^{i-1} p_{\mbox{\scriptsize sol}}(n), & i = j \\
p_{\mbox{\scriptsize sol}}(j), & i>j
\end{array}
\right. 
\label{markovchainequation}
\ee
for $i,j = 0 \ldots M$.
The transition probability $p_{ii}$ can be written in terms of the
hypergeometric function $_2{\cal F}_1(a,b,c;z)$~\cite{arfken} as
\[
p_{ii} =  p_{\mbox{\scriptsize sol}}(i) 
_2{\cal F}_1(1,i-M,i+1;\frac{p_v}{p_v-1})     .
\]
This approximation
ignores the detailed mechanisms of the MC-algorithm 
(which can only change a state locally), instead we consider
transitions between classes of states with the same energy.

\begin{figure}
\centering
\leavevmode
\epsfxsize = .75 \columnwidth
\epsfbox{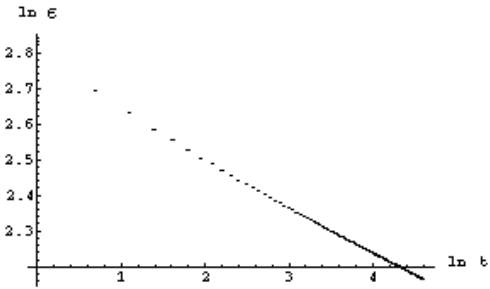}
\caption{Log-log plot of the behavior of
the Markov chain defined by equation~(\ref{markovchainequation}).}
\label{markovrelaxationfigur}
\end{figure}

Using this transition matrix to evolve the states in
time numerically, power law
relaxation behavior is found, see figure~\ref{markovrelaxationfigur}.
The exponent
differs from that in our MC simulations. One reason for this is that
the MC simulations only allow single-spin flips, while
equation~(\ref{markovchainequation}) allows jumps between arbitrary
spin configurations --- the global and the local energy landscapes
are different. We have performed simulations using a MC
algorithm that changes the entire spin configuration instead of just a
single spin. The results indicate a slower power law relaxation, 
with an exponent that is consistent with figure~\ref{markovrelaxationfigur}.
In contrast to the regular MC simulations, the exponent here showed a
weak dependence on system size. The dependence was the same for both
the simulations and time-evolution using
equation~(\ref{markovchainequation}).

In the MC algorithm, $N$ attempts to flip a spin are made in each time step.
Inspired by statistics of the energy landscape, we
have made simulations where we require that there are $N$ 
{\em accepted} spin flips in each time step (in $T=0$
simulations, this means that time is not increased for flip attempts
to higher energy.) Time was increased by $1/N$ for each change in the 
spin configuration.
The power law relaxation behavior did not change 
when this modified algorithm was used. 

For these simulations, we can ignore 
all transitions to states with higher energy. Since a single spin
flip is unlikely to decrease the energy by more than one (see
figure~\ref{frustratedconnectivityfigur}), we can also ignore
transitions to states with energy $j < i-1$, and consider a
two-state system.

To estimate the time needed to go from state $i$ to 
$j=i-1$, we assume that $p(i)/p(i-1)$
can be approximated by
\be
p_{\mbox{\scriptsize sol}}(i)/p_{\mbox{\scriptsize sol}}(i-1) = \frac{p_v}{1-p_v} \frac{M-i+1}{i}
\sim \frac{1}{\epsilon}       .
\label{fractionequation}
\ee
Since the average time needed to pass from state $i$ to state $j=i-1$ is
given by the inverse of the transition probability per unit time, we 
then have
\be
\Delta t =  \frac{p(i) + p(i-1)}{p(i-1)} = 
1 + \frac{p_{\mbox{\scriptsize sol}}(i)}{p_{\mbox{\scriptsize sol}}(i-1)} \sim
1 + \frac{1}{\epsilon}  
\label{deltaTeqn}
\ee
for the time needed to go from state $i$ to $i-1$.

Another viewpoint is to consider the
free energy barriers, $\Delta f$, in the problem. Since 
transitions to states with higher energy are not allowed
by the $T=0$ Monte Carlo dynamics, we can neglect the
energy part of $\Delta f$. The entropic part
will however be positive, 
\be
\Delta f = -k_b T \log{N(j)} + k_b T \log{N(i)} = 
k_b T \log{\frac{p_{\mbox{\scriptsize sol}}(i)}{p_{\mbox{\scriptsize sol}}(j)}}
\label{barriereqn}
\ee
for a transition from $i$ to $j$. The time needed to overcome
this free energy barrier is 
\be
\tau \sim \exp{(\frac{\Delta f}{k_b T})}   \sim
\frac{p_{\mbox{\scriptsize sol}}(i)}{p_{\mbox{\scriptsize sol}}(j)}  \sim \frac{1}{\epsilon}  ,
\label{relaxationequation}
\ee
if $i=j+1$ and $\epsilon = i / N$.

By inverting equations~(\ref{deltaTeqn})
and~(\ref{relaxationequation}), power law relaxation 
with an asymptotic $t^{-1}$ dependence can be obtained.
This is justified for large $\gamma$,
where all free energy barriers have roughly the same size,
but not for $\gamma$ below the freezing transition.
Similar difficulties also arise, e.g., for coarsening, 
where arguments for relaxation usually are given for
nucleation, where a single domain grows or shrinks. 
It is then justifiable
to invert the equation relating the energy of the domain
and the time needed to shrink it. 
In the coalescence regime, however, several such domains form and
grow separately before they coalesce. The inversion is then
not allowed, even though the same relaxation behavior is found
in simulations.

The power law with $t^{-1}$ dependence approximately
matches the behavior of \kcol\ for critically constrained
$\gamma$'s (see table~\ref{colrelaxtable}).
For overconstrained problems, the relaxation is slower than
$t^{-1}$. This could depend on the local
energy landscape being different from the
global when the problems are overconstrained.
The arguments above should be viewed
only as a simplistic first approximation, and cannot 
be expected to reproduce all important features of
the numerical results.

The barriers in equation~(\ref{barriereqn}), 
which are proportional to the 
logarithm of the energy, do not appear in any of the 
coarsening classes introduced by Lai~et~al~\cite{laimazenkovalls}.
The difference seems to be that here there are 
entropy barriers. That the entropy is high for low energies 
is explained by the fact that the 
interactions are antiferromagnetic and the variables are Potts
spins. If each spin can have $K$
different values and the interactions are ferromagnetic,
each satisfied bond can be satisfied in $K$ different ways,
while there are $K^2-K$ choices for the unsatisfied ones.
If the interactions are antiferromagnetic, each
satisfied bond has $K^2-K$ possibilities and the
unsatisfied $K$. Since there are
more satisfied bonds than unsatisfied, an antiferromagnetic
problem is less restricted, at least for $K>2$. 
As always, this reasoning ignores all correlations
between constraints in the problem, such as triangles in
graphs.


\section{Conclusions and Discussion}
\label{discussion}

We have shown that there are additional transitions in
the behavior of two \NPCclass\ problems --- graph
colorability and satisfiability. There is a freezing transition,
and the relaxation changes from exponential to power law
as the difficulty of the problem increases. 
These transitions occur for smaller values
of the parameters $\gamma$ and $\alpha$ than the transitions
in solvability and search cost. 

We also studied the effects of using only colorable graphs,
and measured other quantities, such as the fraction of
persistent spins for \ksat, which also showed a transition.

A simple heuristic argument for the form of the relaxation 
behavior depending on entropy barriers was also given.
The free energy barriers in these
problems have a less pronounced energy dependence
than in other models.
One can compare them to systems where logarithmic
relaxation has been found for $T\neq 0$ and $T=0$ 
MC simulations, such as in~\cite{shoreholzersethna}, 
where a tiling model containing no randomness at all
was studied. The main difference 
between these systems and ours seems to be that they have local 
interactions and frustration, whereas we have infinite-range
interactions and frustration.

The problems with logarithmic relaxation are 
in some ways simpler than the \NPCclass\ problems studied here.
Their ground states can always be found in polynomial time, and
they do not contain any difficult optimization problem. On the other hand,
\ksat\ and \kcol\ have ground states that may
require exponential time to find. Our results show that it can be easier
to find approximate solutions to these hard problems than
for the easy problems giving logarithmic relaxation.
Of course, to actually find the 
ground state with the MC algorithm is very difficult for \kcol\
and \ksat. What the \NPCclass\ problems gain
in finding an approximate solution is lost when 
it comes to finding the ideal, ground state solution.

Other problems where entropy barriers and 
logarithmic relaxation are found include
the backgammon model introduced by Ritort~\cite{ritort} and
the random-field Ising model~\cite{villain,grinsteinfernandez}.
The backgammon model contains no energy barriers at all, only 
entropy barriers, and can be solved~\cite{godrechebouchaudmezard}
by considering random-walk models with entropy barriers. 
In our models, the entropy 
barriers scale as the logarithm of the energy. This means
that the time needed to overcome the free energy barrier loses
its exponential dependence and becomes algebraic, leading to
power law relaxation. We have also found this behavior in
more general constraint satisfaction problems than
\kcol\ and \ksat~\cite{unpublished}. 
Relaxation behavior similar to that in our models
has also been obtained
by Campellone~et~al for a short-range 
$p$-spin glass model~\cite{campellonecoluzziparisi}.
These models should be studied further.

We see several possibilities for future work in this field. 
More
detailed studies of the behavior of the fraction of persistent
spins as well as of 
damage spreading and hysteresis in strongly disordered systems are needed.
Since hard optimization problems are also spin glasses, a search for
spin glass characteristics such as aging and chaotic responses to 
small perturbations in the hamiltonian may also 
be interesting. 
We are preparing a study of the energy landscapes of these and
related models. 
Investigating the connection between the topology 
of the random graph and that of the energy landscape may also provide 
more insight into why \NPCclass\ problems are difficult.

\section*{Acknowledgements} 
We would like to thank an anonymous referee
for useful comments.

\end{document}